\documentclass[12]{article}
\usepackage{url} %this package should fix any errors with URLs in refs.
\usepackage{lineno}

\usepackage[utf8x]{inputenc}
\usepackage{lmodern,textcomp}

\usepackage{soul}

\usepackage{float}

\usepackage[]{todonotes}
\usepackage{graphicx}

\usepackage[left=0.75in, right=0.75in, top=1in]{geometry}

\begin{document}

%%%%%%%%%%%%%%%%%%%%%%%%%%%%%%%%%%%%%%%%%%%%%%%
%  TITLE
%
% (A title should be specific, informative, and brief. Use
% abbreviations only if they are defined in the abstract. Titles that
% start with general keywords then specific terms are optimized in
% searches)
%
%%%%%%%%%%%%%%%%%%%%%%%%%%%%%%%%%%%%%%%%%%%%%%%

% Example: \title{This is a test title}

\title{SoPhAr: Solar Phased-Arrays to boost the range of electric and hydrogen airliners in a solar world}

% The \author macro works with any number of authors. There are two commands
% used to separate the names and addresses of multiple authors: \And and \AND.
%
% Using \And between authors leaves it to LaTeX to determine where to break the
% lines. Using \AND forces a line break at that point. So, if LaTeX puts 3 of 4
% authors names on the first line, and the last on the second line, try using
% \AND instead of \And before the third author name.

\author{Christian Claudel\footnote{Associate Professor, Maseeh department of Civil, Architectural and Environmental Engineering, University of Texas, Austin. 301E E Dean Keeton St C1761, Austin, TX 78712. Email: \texttt{christian.claudel@utexas.edu}}}

\maketitle

%\begin{keypoints}

%\item Salt nanoparticles from ethanol anti-solvent brine precipitation are monodisperse.

%\item Marine cloud brightening (MCB) can utilise Unmanned Aerial Vehicles (UAVs) distributing these salt particles.

%\item This system may be cheaper and has higher Technology Readiness Level (TRL) than alternatives.
%\end{keypoints}

\begin{abstract}
In late 2022, ICAO member states adopted a long-term global aspirational goal (LTAG) to achieve net zero carbon emissions from international aviation by 2050. To date however, no eonomically scalable solution to the aviation decarbonization problem has been proposed. Despite considerable research on potential alternative fuels including e-fuels, Sustainable Aviation Fuel (SAF), Hydrogen or Ammonia, and extensive research on purely electric propulsion, low-carbon propulsion methods are unable to replace fossil-fuels for with identical or better economics.
A possible alternative to current propulsion technologies is to directly beam the required propulsive power to aircraft. Several techniques have been considered to date, in particular laser energy beaming and microwave energy beaming. This paper proposes a possible concept where future airliners are mostly powered with ground-generated power. With expected improvements and scaling in solar panel manufacturing, the proposed concept would be economically competitive even with current jet fuel prices, while considerably reducing CO2 emissions.
\end{abstract}

\section{Introduction}

Commercial aviation heavily relies on fossil fuels for propulsion, and, to date,  technical limitations prevent the deployment of alternative energy sources at scale. Fossil fuels have considerable advantages: they are relatively inexpensive to extract and have considerable energy density, unlike alternate low-carbon energy storage techniques.

One option to reduce fossil fuel usage is to generate synthetic fuels that have similar characteristics as existing jet fuel. Synthetic fuels include SAF~\cite{UNDAVALLI2023100876,WATSON2024141472} (Sustainable Aviation Fuel) or e-fuels made with green hydrogen (generated with solar, hydroelectricity or wind power sources). Synthetic fuels such as SAF are drop-in solutions that does not require significant aircraft or engine redesign, A major challenge of drop-in solutions however is the considerable energy, cost and infrastructure requirements associated with their production and refining, which offsets their cost benefits in terms of aircraft operations. In particular, e-fuels require the construction and maintenance of a large number of hydrogen production and chemical plants, while SAFs require considerable agricultural production for the feedstock, potentially competing with food production (SAF), as well as large construction and maintenance costs for the production facilities. Furthermore, these fuels are expected to be considerably more expensive~\cite{ESWARAN2021111516} than current oil-derived fuels, except for a small fraction~\cite{doi:10.1021/acssuschemeng.3c02147} that could be cost-effective, albeit this fraction could only replace 10-25\% of total current fuel usage.

Another option is to investigate new types of fuel that are potentially easier and more efficiently produced than SAF or e-fuels. Possible energy storage media include Hydrogen, Ammonia~\cite{doi:10.1021/acsenergylett.1c02189}, or batteries, though all of these options are associated with relatively poor energy densities (either per unit volume, or per unit mass) in comparison with jet fuel, e-fuel or SAF. This low energy density would in turn not only severely limit the range~\cite{gao2022hydrogen,sampson2023feasibility} of such aircraft, but it would also require considerable aircraft redesign to accommodate larger tanks~\cite{degirmenci2023challenges} or batteries~\cite{viswanathan2022challenges}, potentially severely restricting payload. Furthermore, numerous safety problems are currently unresolved~\cite{viswanathan2022challenges}, in particular battery fires, ammonia toxicity in case of severe crash, and post-crash fires for Hydrogen.

A third option is to power future airliners remotely~\cite{orndorff2023gradient}, using energy beaming~\cite{rodenbeck2021microwave}. Energy beaming involves remotely sending power required for propulsion to an aircraft. While these methods have been tested on the ground~\cite{jaffe2022spooky} and in UAVs~\cite{rodenbeck2021microwave}, to date, no piloted aircraft has been flown under beam-power. Considerable interest in this area exist however, in particular a concept of operation for beam-powered e-VTOL~\cite{sheth2023energy,sheth2023concept} has recently been proposed, though the proposed beaming power levels are in the orders of tens of kW, which is orders of magnitude less than the power required by larger commercial aircraft. 

Advantages of beam-powered aircraft include potentially high end-to-end efficiencies, and the possibility of reducing aircraft weight, since the power source would remain on the ground. However, major drawbacks exist, in particular the physical impossibility of beaming energy over large distances unless antennas of considerable dimensions are considered, in the case of microwave energy beaming. Because of their much shorter wavelength, visible or infrared lasers keep very good directivity over the required ranges (tens of km), though they pose unacceptable safety risks to the population and to pilots, when used at power levels required by commercial aviation. Furthermore, unlike microwave beaming, laser beaming is not an all-weather solution. 

\section{Aviation in the context of the energy transition}

\begin{figure}
    \centering
    \includegraphics[width=1\linewidth]{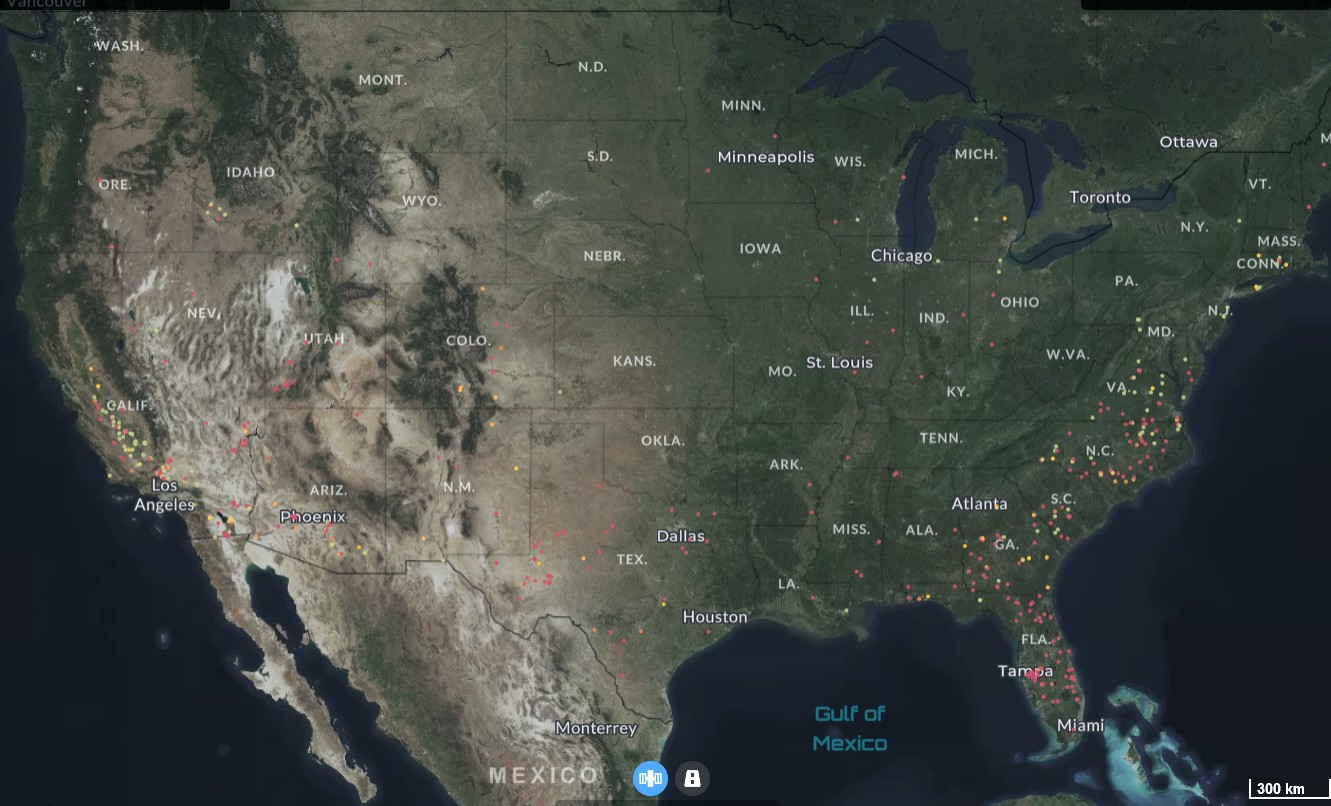}
    \caption{Locations of solar farms with capacity exceeding 20MW as of November 2023 (Source: USGS)}
    \label{f:2}
\end{figure}

Large scale photovoltaic facilities have recently increased in popularity due to lower lifecycle costs and CO2 emissions than competing sources. Figure~\ref{f:2} shows the map of large capacity solar power plants in the US as of November 2023. 

With increasing global warming concerns and dwindling fossil fuel reserves, the world is expected to add considerable wind and solar power generation capacity~\cite{grossmann2012investment} by 2050. By this date, approximately half of the US generated electricity will be solar~\cite{JAYADEV2020114267} or wind-derived, with some scenarios envisioning near complete renewable energy production~\cite{jacobson2022low}.

The increased solar power production will require a significant area to be dedicated to solar panel deployment.Indeed, to meet the near$50\%$ solar power generation objectives, a solar power capacity on the order of 1600 GW is required~\cite{linkenergy}, requiring the deployment of solar panels over $0.5\%$ of the United States. 

A large fraction of these solar panels will be deployed in large-scale solar farms. Solar farms offer considerable advantages over rooftop systems, in particular the ease of maintenance, control, and the economies of scales associated with concentrating solar panels and associated equipment at the same location. Conservatively assuming that only $0.1\%$ of the United States will be covered by solar farms by 2050, this would translate into approximately 8,000 solar farms of $1 km^2$ area each, that is, one $km^2$ solar farm every $1,000km^2$ in average (or equivalently one solar farm every about $30 km$).

\section{Energy beamimg: solar farms as phased arrays}

\begin{figure}
    \centering
    \includegraphics[width=1\linewidth]{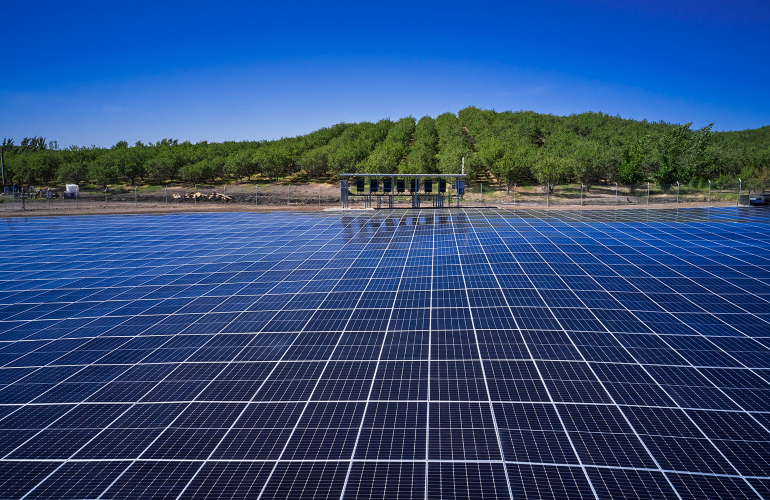}
    \caption{Ground mounted 100 MW solar farm (Credit: Erthos)}
    \label{f:1}
\end{figure}

For the proposed concept, we assume that the solar farm is made by mounting solar panels horizontally on a flat surface, though it could be extended to more complex geometries. As of 2024, this type of deployment is becoming increasingly popular, due to faster build time, wind resistance and decreased capital and maintenance, see for example Figure~\ref{f:1}. One disadvantage of ground-mounted panels is the need for higher required surface area of panels, though the effect is minor at low latitudes.

Recently, some authors have considered printing antenna arrays onto solar panels, see for example~\cite{yekan2017conformal,peter2014novel,bermudez2023optically,li2019silver}, or the recent ARACHNE project~\cite{rodenbeck2021microwave}, with the SSPRITE subsystem, currently under development, that will test integrated (sandwiched) solar tiles comprising photovoltaics, RF and support electronics.

\subsection{Phased arrays}

Using these advances, microwave emitters can be integrated with solar panels as part of the manufacturing process. Since the cost of the antennas is relatively low, and their power production penalty is similarly relatively low (with very high transparency), a large portion of solar panels could be built with these RF emitting layers with relatively low additional cost and similarly favorable power production economics.

\begin{figure}
    \centering
    \includegraphics[width=1\linewidth]{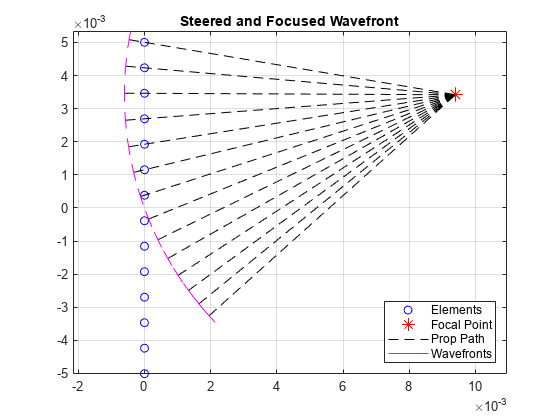}
    \caption{Simulation of wave propagation from a phased array, focusing at a particular (controllable) point in space (source: Mathworks)}
    \label{f:3}
\end{figure}

These dual solar/RF tiles will form the backbone of the proposed system, an horizontal, ground mounted phased array of independently controllable solar/RF panels mounted side-by side over extended areas. In the below calculations, we assume that the solar farm has a diameter on the order of 1 km, 

Leveraging the entire solar farm as a large aperture, the system has the capability to direct and focus energy at controllable points in space~\cite{mailloux1982phased}, using beamforming techniques. Essentially, phase shifters modify the phases of individual elements in such a way the that waves they emit constructively interfere at a particular point in space (the targeted point), and destructively interfere at other points. An illustration of this effect is shown in Figure~\ref{f:3}.

\begin{figure}
    \centering
    \includegraphics[width=1\linewidth]{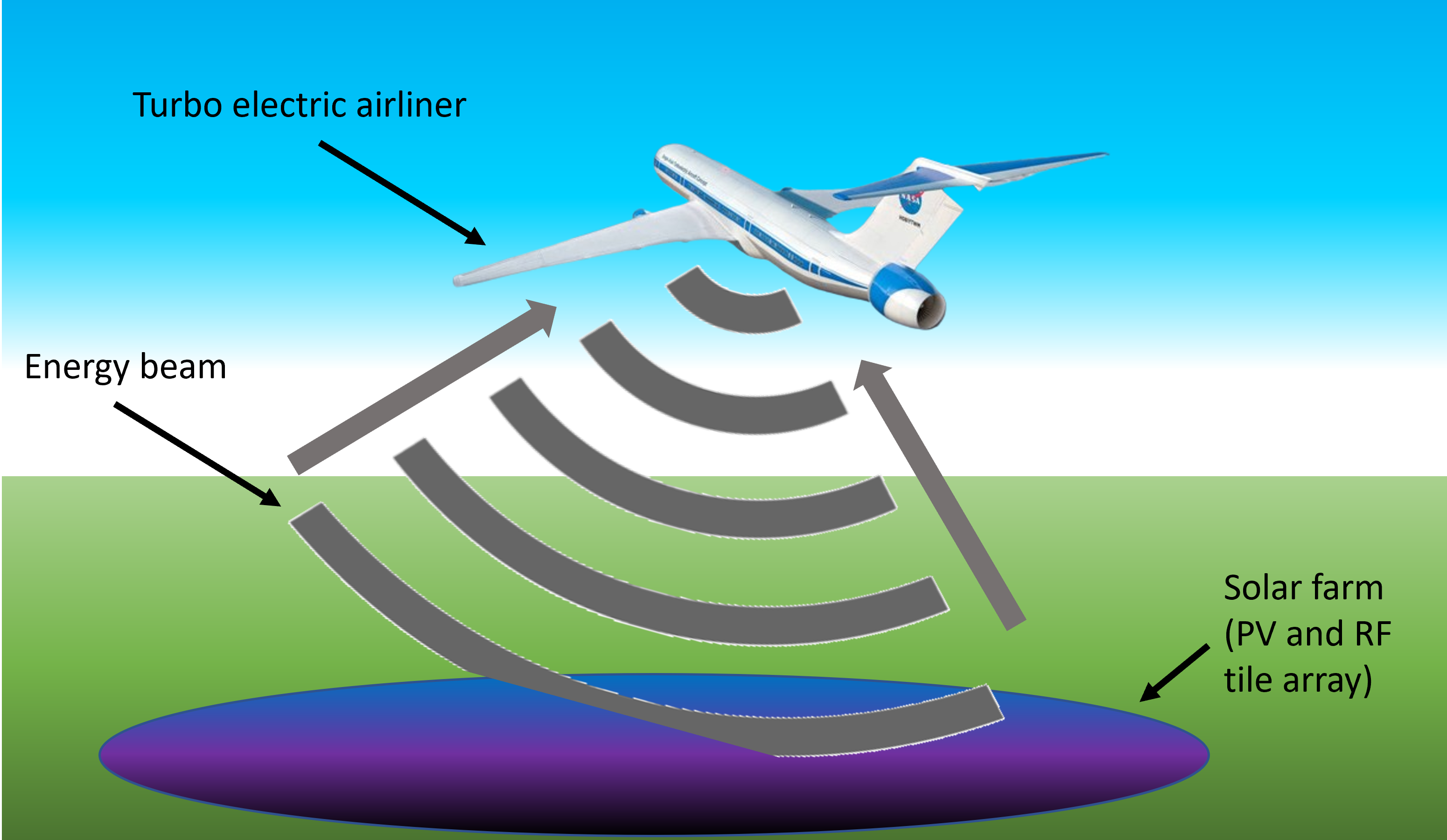}
    \caption{Illustration of energy beaming from an array of PV-RF tiles using beamforming. Energy is beamed to a turbo electric airliner.}
    \label{f:4}
\end{figure}

The minimum focusing angle $\theta$ is diffraction limited: $\theta=1.22\cdot \frac{\lambda}{D}$, where $\lambda$ is the wavelength emitted by the system, and $D$ is the aperture of the system. With a 1 km aperture and a wavelength of $10 cm$ (S band), the system can focus $87\%$ of the radiated energy into a disk of diameter $1.22 m$ at 10,000 meters distance (cruising altitude of an airliner), and nearly $100 \%$ of the radiated energy into a disk of diameter $3.7 m$. Figure~\ref{f:4} illustrates the concept with a turbo electric airliner.

Some development of the solar/RF panel technology is required to make this economically viable and scalable. This development can leverage existing efforts for SSP (Space Solar Power), which have similar beam control and economic scalability objectives. In particular, the future work on SSPRITE can be used both for satellite (SSP) and the present terrestrial application.

\subsection{Physical considerations}

The array of solar panel/RF tiles should be deployed in a relatively planar surface, though it can tolerate some deviations (as in conformal antenna arrays~\cite{peng2021conformal}.

In order to maintain efficiency, the array needs to satisfy some design constraints:

\begin{itemize}
\item Due to the Thinned Array Curse~\cite{jackson2020tile}, the surface area of the solar farm should be mostly covered with solar/RF sandwiched tiles. In the present case, this means that the surface area of the gaps between solar panels should be a small fraction of the total surface area.
\item Choice of the used wavelength: the spacing between emitting elements depends on the chosen wavelength, and define the ability to steer the beam in different directions to power the plane as it flies around the array. In order to steer the beam over $90^{\circ}$ (which allows the farm to power an aircraft flying at 10 km altitude within a 10 km radius of the farm), a spacing of $0.5 \lambda$ is required for each emitting element (thus, each element would be a $5cm\times 5cm$ square a $3 GHz$ frequency). Note that lower frequencies would scale more easily, since the elements would be comparatively larger, with less required electronics and emitters. For example, using a frequency of $1 GHz$, each element would be a $15cm \times 15 cm$ square, which would be less expensive to manufacture. Longer wavelengths are associated with lower costs, though they also have poorer focusing capabilities: a $1GHz$ source could focus energy in a $3m$ diameter disk at 10,000 m distance, while a $3GHz$ source could focus the energy down to a $1m$ diameter disk. Based on the respective physical sizes of aircraft and solar farms, the minimum usable frequency (which would lead to minimal costs) would be on the order of $1 GHz$.
\end{itemize}

\section{Safety and economics}

\subsection{Safety}

One of the major advantages of this system is the distributed nature of the emitters. Since the input beamed power is at most $100 MW$ (resulting in about $20 MW$ net power for the aircraft, at $20\%$ overall emission-conversion to DC efficiency), power densities at the surface of the solar farm are less than $100 W/m^2$, considerably safer than for instance SSP applications ($1 kW/m^2$), with likely minor impact to the wildlife. Energy levels are only very high in close proximity of the targeted aircraft. Aircraft will thus need to shield microwave radiation from the passengers, in case of accidental steering of the beam on windows or other sensitive areas. RF opaque windows can be conceived to shield pilots and passengers from this potential accidental exposure. 

Reflections do not cause danger to people in the ground. Assuming that $100\%$ of a $100 MW$ beam reflected out from a planar surface on the airplane (worst-case scenario), at $10 km$ distance, the power densities on the ground will not exceed $100 W/m^2$ during the brief exposure transient.

\subsection{Economics}
The economics of this system depend on the overall transmission efficiency, the availability of this system, and its overall lifecycle costs. The lifecycle cost is difficult to evaluate since the technology of RF/solar cells is at its infancy, though the major components of the system (oscillators, phase shifters, antenna arrays) will probably cost a fraction of the overall solar panel cost at scale (this ratio depends on the chosen solar panel technology). To estimate this fraction,~\cite{jackson2020tile} anticipates added costs (at scale) of $\$100/m^2$, which correspond to a $50\%$ increase over the current typical cost of solar panels ($\$200/m^2$ or $\$1/W$). 

This system can be available at all times. Microwave power can be efficiently transmitted through clouds and rain, and, if the solar farm does not generate sufficient power for beaming (night, bad weather), grid power can be directly used to power emitters tiles, turning the solar farm into a load for the power grid. The system can also (through beamforming techniques) supply powers at multiple aircraft at any given time. Total beamable power is limited mainly by grid current limits and power constraints.

Deployment locations and schedule should be chosen to maximize benefits for energy generation and energy beaming purposes. Solar farms can be deployed in unused land around airports to power aircraft in holding pattern and during climb. The system can then be expanded around main aircraft routes from and to major airports, and (as floating solar farms) can be also installed off-shore (close to population centers) near oceanic routes and in the great lakes, to expand coverage.

Power generated through this system will be competitive with alternate aircraft energy sources, since solar power is already one of the most inexpensive source of energy (which can reach as low as $\$24/MWh$ as of 2024, and lower in the coming decades). Assuming a $50\%$ increase of the cost per MWh due to the additional RF layers, electronics and emitters (which increase production cost), we can assume an at-scale cost of $\$36/MWh$ in the future from this system, and probably less given improvements in solar panel production capacities, improvements in material sciences for the emission components, and economies of scale. 

To evaluate the competitiveness of this system, we consider an hypothetical turbo-electric~\cite{sayed2021review} Airbus A320 with a mass of $50,000 kg$, a fuel burn of $2,400 kg/h$ (turbofan only), a cruise lift to drag-ratio of 18, and a combined motor/propulsive efficiency of 0.6 for the electric fan. The total cost of fuel is $\$1992/h$ as of March 2024. This A320 needs $11.3 MW$ to cruise at $250 m/s$. At the production and beaming cost of $\$36/MWh$, the system is competitive with current jet fuel cost whenever end-to-end efficiency exceeds $20\%$. This level of efficiency is achievable~\cite{rodenbeck2021microwave} as of today, with end-to-end beaming efficiencies up to $54\%$ demonstrated in some tests.

For longer flights, savings are even larger since the turbo electric airliner does not have to carry as much fuel, reducing its required propulsive power. For a flight above the continental US, assuming solar power deployment scenarios for 2020-2050, most of the flight would occur in range of one of these solar farm arrays, and fuel burn would be a minor fraction of the total propulsion energy.

\section{Power reception}

\begin{figure}
    \centering
    \includegraphics[width=1\linewidth]{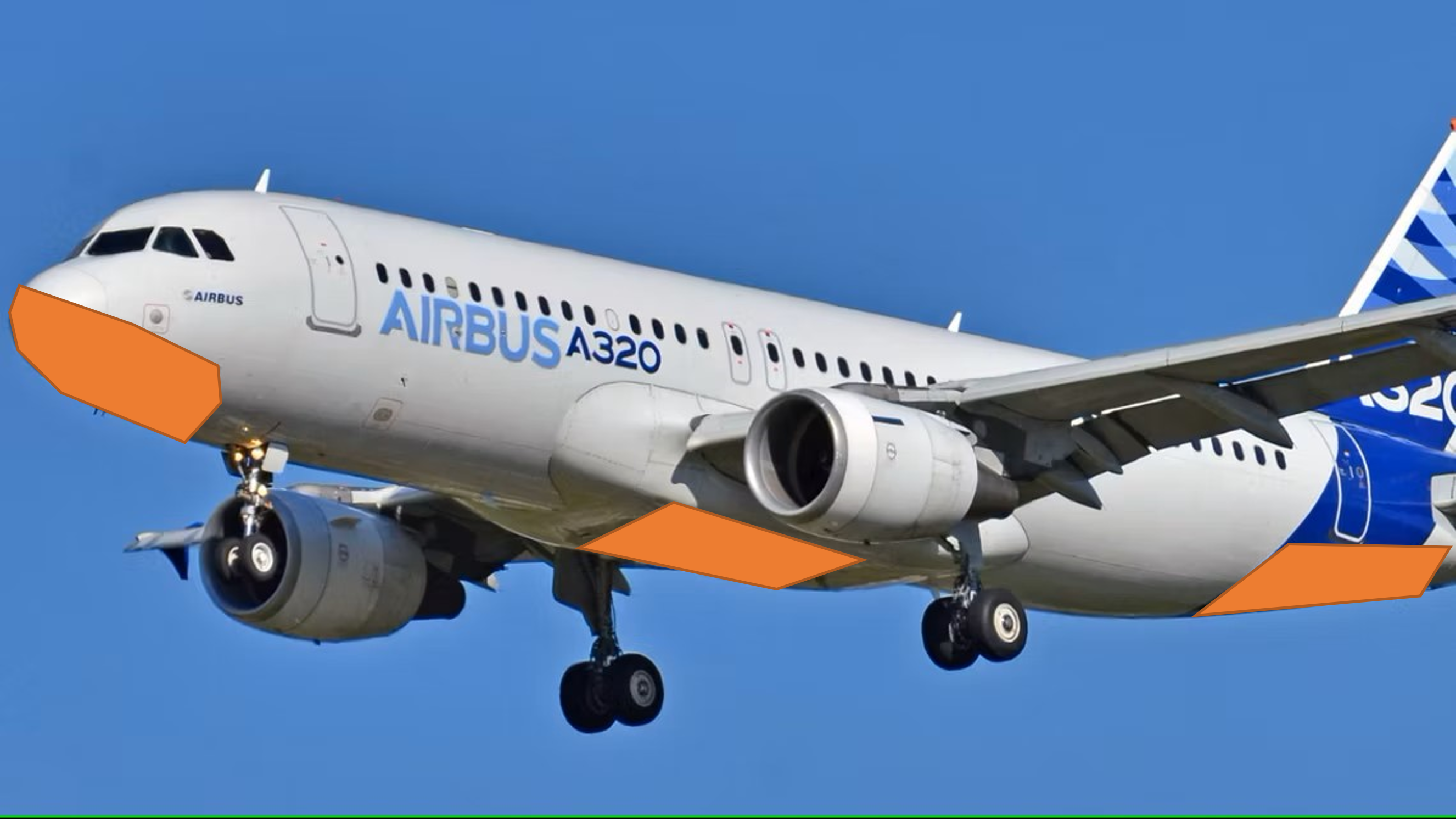}
    \caption{Potential receiver locations for an Airbus A320 (orange areas).}
    \label{f:5}
\end{figure}

This system relies on energy beaming to provide a large fraction of the energy used by an airliner. This power is received and converted into usable DC power with rectenna arrays. While the conversion process is outside of the scope of this article, high efficiencies of $85\%$ have been achieved for reception~\cite{sheth2023concept}. To minimize losses and make the system economically scalable, it is essential to be able to send power from 15-20 km away, which would provide coverage virtually anywhere in the continental United States. Airplane receivers arrays should thus be mounted to minimize reflections losses by ensuring that incidence angles remain low. Some regions of an airplane are most favorable: underside for overhead flying aircraft, lower front fuselage for incoming aircraft and lower tail section for outgoing aircraft. Depending on the particular path of the aircraft, power could be beamed on one or multiple of these receivers to minimize cosine losses. Multiple arrays of these receivers should be installed on aircraft to maximize efficiency and improve redundancy (Figure~\ref{f:5}).

\section{Conclusion}

This paper discusses a possible concept for energy beaming in the age of electrified aircraft propulsion (turbo-electric, battery-electric or fuel-cell). Using recent advances in the integration of RF antennas with solar panels, each future solar farm can become a phased array emitter, that can efficiently beam energy (thanks to its very large aperture) to airliners flying around it. With an expected 1600 GW of solar capacity by 2050, a very large fraction of the US (and world) land surface area will be covered by these farms, enabling future airliners to greatly reduce fuel burn, or equivalently boost the range of hydrogen or battery-electric aircraft. The economics of this system are favorable as of 2024, and the gap will widen in the future given ever increasing fuel costs and decreasing solar power prices.

\section{Acknowledgements}

The author would like to thank Peng Wei, Julien Barthes, Fabian Hoffmann, Atif Shamim, Paul Jaffe, Elias Wilcoski, Seth Schisler, Les Johnson and Nikolai Joseph for fruitful conversations and discussions leading to this paper.

%\subsection{Location optimization}
%
%The use of UAVs to deliver salt nanoparticles bears the potential for optimizing the fraction of low clouds that are sprayed. Since UAVs can be steered to changing locations relatively quickly, spraying of nanoparticles in areas without the potential for substantial brightening (e.g., cloud-free regions, highly polluted clouds, very deep clouds) can be avoided. Thus, by not spraying all maritime regions of the globe but only specific regions, e.g., the stable stratocumulus decks over the eastern subtropical oceans off the coasts of California, Chile, and Namibia, one can increase the spatially and temporally averaged low cloud cover considerably from $0.33$ to $0.68$ \cite{wood2021assessing}. Following the heuristic arguments of \cite{wood2021assessing}, a reduction in the fraction of sprayed maritime regions from $1.0$ to, say, $0.2$, would result in an increase in the low cloud cover of the ideally sprayed regions from $0.33$ to $0.46$. As the product of these two quantities is proportional to the MCB radiative forcing, a fifth of the global global sprayer fleet can still cause one third of the radiative forcing. As the use of UAVs would also allow to adapt the spraying to seasonal and synoptical changes, potentially guided by satellite observations, an even higher radiative forcing is conceivable by targeting the most susceptible regions for spraying only, thereby reducing the mass of sprayed NaCl and the number of required UAVs even further. Nonetheless, more research is required to assess these additional benefits of UAVs. 

\clearpage
\bibliographystyle{abbrv}
\bibliography{refpaper}

%%%%%%%%%%%%%%%%%%%%%%%%%%%%%%%%%%%%%%%%%%%%%%%%%%%%%%%%%%%%

\appendix

\end{document}